\documentclass[prb,twocolumn,amsmath,amssymb,amsbsy,superscriptaddress,floatfix,nobalancelastpage]{revtex4-1}

\usepackage{graphicx,bm,times}
\usepackage{color}
\usepackage[]{natbib}

\begin{document}

\title{A de Haas van Alphen study of the role of 4f electrons in antiferromagnetic CeZn$_{11}$ as compared to its non-magnetic analogue LaZn$_{11}$}

\author{S. F. Blake}
\affiliation{Clarendon Laboratory, Department of Physics, University of Oxford, Parks Road, Oxford OX1 3PU, UK}

\author{H. Hodovanets}
\affiliation{Ames Laboratory, Iowa State University, Ames, Iowa 50011, USA}
\affiliation{Department of Physics and Astronomy, Iowa State University, Ames, Iowa 50011, USA}

\author{A. McCollam}
\affiliation{High Field Magnet Laboratory (HFML-EMFL), Radboud University, 6525 ED
Nijmegen, Netherlands}

\author{S. L. Bud'ko}
\affiliation{Ames Laboratory, Iowa State University, Ames, Iowa 50011, USA}
\affiliation{Department of Physics and Astronomy, Iowa State University, Ames, Iowa 50011, USA}

\author{P. C. Canfield}
\affiliation{Ames Laboratory, Iowa State University, Ames, Iowa 50011, USA}
\affiliation{Department of Physics and Astronomy, Iowa State University, Ames, Iowa 50011, USA}

\author{A. I. Coldea}
\affiliation{Clarendon Laboratory, Department of Physics,
University of Oxford, Parks Road, Oxford OX1 3PU, UK}

\begin{abstract}
We present a de Haas-van Alphen study of the Fermi surface of the low temperature antiferromagnet CeZn$_{11}$ and its non-magnetic analogue LaZn$_{11}$, measured by torque magnetometry up to fields of $33\,\mathrm{T}$ and at temperatures down to $320\,\mathrm{mK}$. Both systems possess similar de Haas-van Alphen frequencies, with three clear sets of features -- ranging from $50\,\mathrm{T}$ to $4\,\mathrm{kT}$ -- corresponding to three bands of a complex Fermi surface, with an expected fourth band also seen weakly in CeZn$_{11}$. The effective masses of the charge carriers are very light ($<1\,m_e$) in LaZn$_{11}$ but a factor of $2$ - $4$ larger in CeZn$_{11}$, indicative of stronger electronic correlations. We perform detailed density functional theory (DFT) calculations for CeZn$_{11}$ and find that only DFT+$U$ calculations with $U$=1.5\,$\mathrm{eV}$,
 which localize the $4f$ states, provide a good match to the measured de Haas-van Alphen frequencies, once the presence of magnetic breakdown orbits is also considered. Our study suggests that the Fermi surface of CeZn$_{11}$ is very close to that of LaZn$_{11}$ being dominated by Zn $3d$, as the Ce $4f$ states are localised and have little influence on its electronic structure, however, they are responsible for its magnetic order and contribute to enhance electronic correlations.
\end{abstract}

\date{\today}
\maketitle

\section{Introduction}

Intermetallic compounds containing Ce have been widely studied for their unconventional electronic and magnetic properties, associated with the heavy fermion behaviour that may arise from interactions between the conduction and the $f$ electrons \cite{Coleman2007}. In these materials the competition between the RKKY interaction \cite{Ruderman1954} -- which favours localisation of the $f$ electrons and an antiferromagnetic (AFM) ground state -- and the Kondo effect \cite{Kasuya1956} -- which favours hybridisation with the conduction electrons and a paramagnetic (PM) ground state -- can lead to regions of unconventional superconductivity \cite{Steglich1979} and non-Fermi liquid behaviour \cite{VanLohneysen1994}, reached by tuning with doping \cite{VanLohneysen1994}, pressure \cite{Sullow1999} or magnetic field \cite{Custers2003} and accompanied by greatly enhanced effective masses and Sommerfeld coefficients \cite{Steglich1979,VanLohneysen1994,Sullow1999,Custers2003}.

As an intermetallic containing Ce, the compound CeZn$_{11}$ -- which crystallises in a tetragonal structure (space group $I4_1 / amd$) with each Ce atom enclosed by a polyhedra of 22 Zn atoms \cite{OZelinskaMConrad2004} -- is therefore of interest as a potential heavy fermion material. Its \textit{caged} crystal structure is similar to the heavy fermion antiferromagnets U$_2$Zn$_{17}$ \cite{Ott1984} and UCd$_{11}$ \cite{Cornelius1999}, which both possess an Actinide surrounded by a polyhedra of Zn or Cd atoms. However the isoelectronic caged compound CeCd$_{11}$ has AFM ordering but no heavy fermion behaviour, with the RKKY interaction dominating and the occupied $f$ states well localised \cite{Yoshiuchi2010}.

CeZn$_{11}$ shows a low temperature AFM state below $T_N = 2 \, \mathrm{K}$, with the saturated magnetic susceptibility and magnetic entropy taking values close to those expected for a Ce$^{3+}$ ion in a $4f^1$ configuration, confirming the magnetic state arises from localised moments at the Ce sites \cite{Hodovanets2013a,Nakazawa1993}. Applying a magnetic field suppresses the AFM phase by $5\,\mathrm{T}$ (for ${\bf B} \parallel [1\,1\,0]$) and the Sommerfeld coefficient, $\gamma$, becomes significantly enhanced around this quantum phase transition, increasing up to $200 \,\mathrm{mJ}/\mathrm{mol}\,\mathrm{K}^2$ before dropping to a smaller value in the PM phase \cite{Hodovanets2013a} -- unlike in the heavy fermion antiferromagnets where it remains strongly enhanced \cite{Ott1984,Cornelius1999}. LaZn$_{11}$, with a similar crystallographic structure but lacking the $4f$ electron, is a non-magnetic analogue to CeZn$_{11}$ and shows a Sommerfeld coefficient of $\gamma = 10.3 \, \mathrm{mJ}/\mathrm{mol}\,\mathrm{K}^2$ \cite{Hodovanets2013a}, consistent with what is expected for a normal metal without correlations. Thus comparing these two systems can help establish the role of the $4f$ electron in determining the electronic and magnetic properties of CeZn$_{11}$.

In this paper we present a study of de Haas-van Alphen (dHvA) oscillations in single crystals of CeZn$_{11}$ and LaZn$_{11}$, measured using torque magnetometry at temperatures down to $320\,\mathrm{mK}$ and magnetic fields up to $33\,\mathrm{T}$. The cyclotron effective masses are extracted from the temperature dependence of the dHvA oscillations, and the experimentally determined frequencies are compared to those predicted by density functional theory (DFT), spin polarised DFT and DFT+$U$ calculations. We find that, with the consideration of magnetic breakdown orbits, the Fermi surface of CeZn$_{11}$ resembles that of LaZn$_{11}$ with the $4f$ electrons localised well below the Fermi level. The effective masses are very light in LaZn$_{11}$,
in close agreement with band structure calculations, and we detect significant mass enhancement caused by
stronger electronic correlations in CeZn$_{11}$.

\section{Experimental details}
The torque experienced by a sample in a magnetic field (${\bf B}$) is given by ${\bm \tau} = {\bf M} \times {\bf B}$ and so acts as an indirect measure of the sample's magnetisation (${\bf M}$), depending on the component perpendicular to the applied field. Therefore, with the exception of entirely isotropic systems for which the torque is zero, torque magnetometry provides a measure of the anisotropic magnetic properties of a system and a way of detecting magnetic phase transitions. Low noise torque magnetometry measurements conducted using piezoresistive cantilevers are also able to measure the dHvA effect, oscillations in the magnetisation of a sample whose frequencies ($F$) in inverse field are related to the size of extremal orbits of the Fermi surface ($A_F$) by the Onsager relation $F= (\hbar / 2 \pi e) \, A_F$, as detailed in Ref.\onlinecite{Onsager1952}. The amplitude of dHvA oscillations additionally contains information about the effective masses and scattering processes of the charge carriers in the system \cite{Shoenberg1984a}.

High quality CeZn$_{11}$ and LaZn$_{11}$ single crystals were grown by the method detailed in \cite{Hodovanets2013a}, with reported residual resistivity ratios in excess of $300$. For torque measurements small, flake-like samples, no bigger than $100 \times 100 \times 20 \, \mu\mathrm{m}$ were selected to limit the deviation experienced by the cantilevers, were cleaved from larger single crystals. The orientations of the flat faces of these samples were determined by single crystal x-ray diffraction. The torque measurements were carried out using Seiko PRC-400 and PRC-120 self-sensing microresistive cantilevers, with the sample affixed by Apiezon N grease so that it may be accurately positioned by hand and its orientation was compared with the X-ray diffraction pictures to be along a high symmetry axis. LaZn$_{11}$ was measured at fields up to $14\,\mathrm{T}$ and temperatures down to $2\,\mathrm{K}$ in a Quantum Design PPMS fitted with a horizontal rotator with a high angular accuracy of less than $0.05^\mathrm{o}$. CeZn$_{11}$ was measured at fields up to $18\,\mathrm{T}$ and temperatures down to $1.4\,\mathrm{K}$ in an Oxford Instruments superconducting magnet and fields up to $33\,\mathrm{T}$ and temperatures down to $0.32\,\mathrm{K}$ in a Bitter resisitive magnet, fitted with a Helium-3 insert, at HFML Nijmegen. In both cases the sample was mounted on a single axis Swedish rotator where the rotation is controlled manually by turning a dial at the top of the probe (accurate to within $1^\mathrm{o}$). DFT calculations were conducted using the Wien2k package \cite{Blaha2001} on a $46\times46\times46$ ($\sim$10,000 $k$-point) grid, with the GGA form of the exchange-correlation energy and spin-orbit coupling included.  Experimental lattice parameters were taken from Ref.~\onlinecite{Hodovanets2013a}. DFT+$U$ calculations were conducted for $U$=0, 1 and 1.5\,$\mathrm{eV}$, with the Hubbard $U$ term applied only to the Ce $4f$ electron.

\section{de Haas van Alphen effect in torque measurements}
\begin{figure}[h]
\centering
\hspace{-2mm}
    \includegraphics[width=0.48\textwidth]{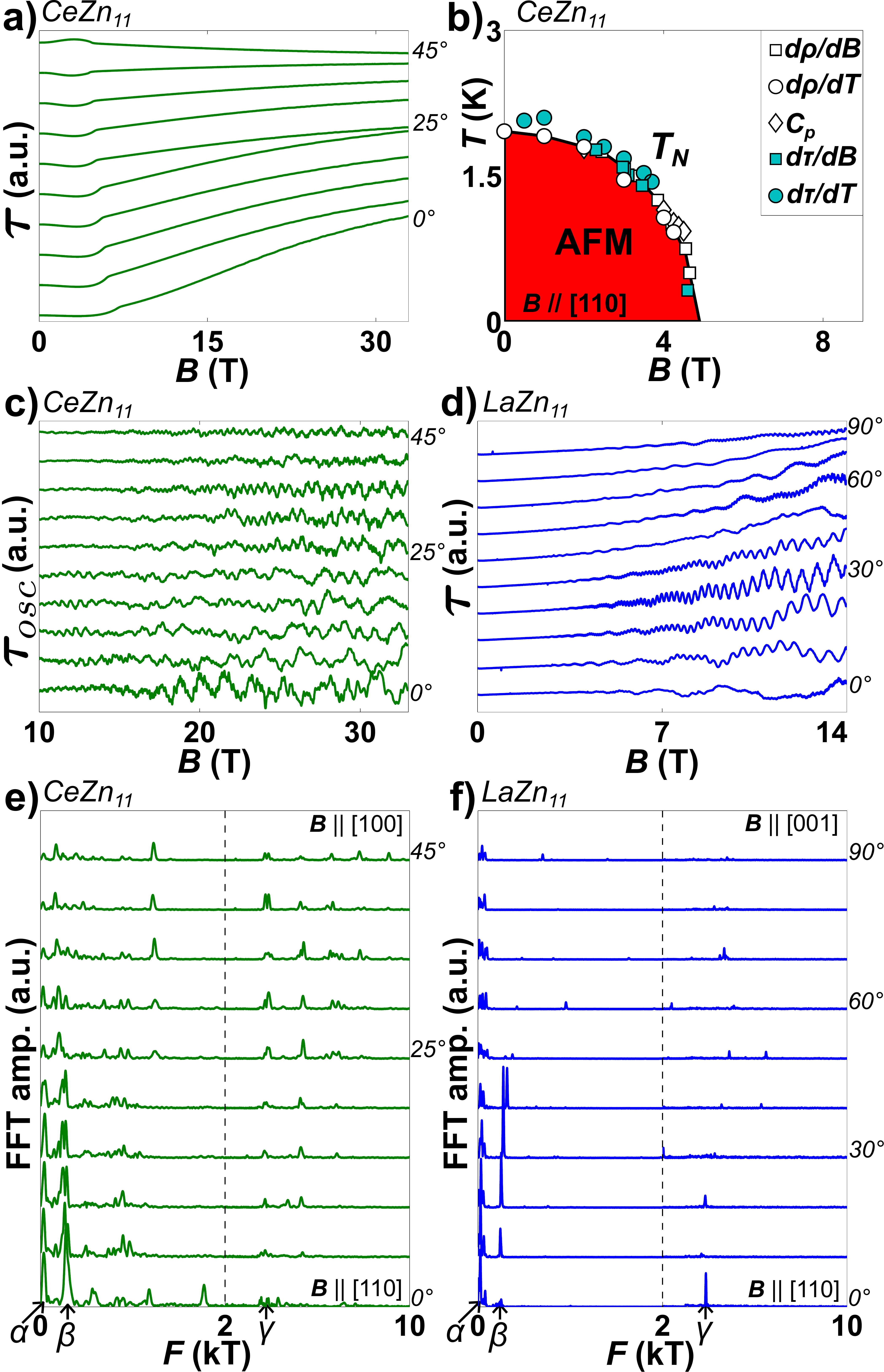}
  \caption{{\bf dHvA oscillations in CeZn$_{11}$ and LaZn$_{11}$.} {\bf CeZn$_{11}$}: {\bf a)} Field dependence of torque at $320\,\mathrm{mK}$ for angles between ${\bf B} \parallel [1\,1\,0]$ ($0^\mathrm{o}$) and ${\bf B} \parallel [1\,0\,0]$ ($45^\mathrm{o}$), alongside {\bf b)} the temperature-field phase diagram at ${\bf B} \parallel [1\,1\,0]$ (white data points taken from \cite{Hodovanets2013a}), {\bf c)} the oscillatory component of torque and {\bf e)} FFTs of the oscillatory component. {\bf LaZn$_{11}$}: {\bf d)} Field dependence of torque at $2\,\mathrm{K}$ and {\bf f)} FFTs of the oscillatory component for angles between ${\bf B} \parallel [1\,1\,0]$ ($0^\mathrm{o}$) and ${\bf B} \parallel [0\,0\,1]$ ($90^\mathrm{o}$). FFTs are performed over a wide field window of $\Delta B = 10$ - $33\,\mathrm{T}$ to distinguish the low frequencies below $F=2\,\mathrm{kT}$ and a narrow high field window of $\Delta B = 20$ - $33\,\mathrm{T}$ to amplify the high frequencies above separated by the vertical dashed line.}
\label{RawData}
\end{figure}

Figure \ref{RawData}a) shows the magnetic field dependence of torque for CeZn$_{11}$, measured up to $33\,\mathrm{T}$ at $320\,\mathrm{mK}$, for fixed field orientations between the different crystallographic planes of the sample: ${\bf B} \parallel [1\,1\,0]$ ($0^\mathrm{o}$), the easy axis of magnetisation \cite{Hodovanets2013a}, and ${\bf B} \parallel [1\,0\,0]$ ($45^\mathrm{o}$). We observe strong kinks in the torque at low fields, corresponding to the magnetic field-induced transition from an AFM to a PM phase, that vary with field orientation from $3\,\mathrm{T}$ at ${\bf B} \parallel [1\,0\,0]$ to $5\,\mathrm{T}$ at ${\bf B} \parallel [1\,1\,0]$. Figure \ref{RawData}b) shows the temperature-field phase diagram of this phase transition along the easy axis ${\bf B} \parallel [1\,1\,0]$, with $T_N=2\,\mathrm{K}$ at zero field and the AFM phase suppressed entirely by $5\,\mathrm{T}$. Our values obtained from torque measurement (taken as the peak in $d \tau / dB$ and $d \tau / d T$) are in good agreement with those previously measured in resistivity ($\rho$) and specific heat ($C_p$) \cite{Hodovanets2013a}, also plotted in Figure \ref{RawData}b).

At all orientations the torque has a strong magnetic background, largest along the easy axis ${\bf B} \parallel [1\,1\,0]$, arising from the localised $4f$ moments of the Ce atom. On top of this background de Haas-van Alphen oscillations are observed, and can be seen clearly at higher fields once the background (approximated to a third order polynomial fit) is subtracted, as shown in Figure \ref{RawData}c). Oscillations are seen across the angular range with considerable variation in both frequency and amplitude, indicative of a complex, three-dimensional Fermi surface. Fast Fourier transforms (FFTs) of the oscillatory component of the torque, shown in Figure \ref{RawData}e), reveal three clear sets of features close to ${\bf B} \parallel [1\,1\,0]$:
  a small frequency
around $50\,\mathrm{T}$ labelled $\alpha$,
a pair of slightly larger frequencies close to $280\,\mathrm{T}$ labelled $\beta$, and a clear frequency at $3.9\,\mathrm{kT}$ labelled $\gamma$. The features $\alpha$ and $\beta$ decrease in both amplitude and frequency at higher angles with rotation away from ${\bf B} \parallel [1\,1\,0]$, but $\gamma$ exists across the entire measured angular range. Additionally there are many unassigned, smaller amplitude frequencies up to $10\,\mathrm{kT}$ that are present at some orientations.

The magnetic field dependence of torque was also measured for LaZn$_{11}$ for field orientations between the crystallographic planes ${\bf B} \parallel [1\,1\,0]$ ($0^\mathrm{o}$) and ${\bf B} \parallel [0\,0\,1]$ ($90^\mathrm{o}$). Figure \ref{RawData}d) shows the torque measurements for LaZn$_{11}$ with dHvA oscillations clearly visible on top of a weak paramagnetic background -- as La does not possess a $4f$ electron the background torque is much smaller than in CeZn$_{11}$. FFTs of the oscillatory component of the torque are shown in Figure \ref{RawData}f), close to ${\bf B} \parallel [1\,1\,0]$ similar features are seen to those in CeZn$_{11}$, suggesting a similarity of the Fermi surface:   a cluster of very small frequencies  labelled $\alpha$, a pair of stronger frequencies around $250\,\mathrm{T}$ labelled $\beta$, and a clear high frequency of $3.9\,\mathrm{kT}$ labelled $\gamma$. Rotating towards ${\bf B} \parallel [0\,0\,1]$ -- a different rotation to that performed for CeZn$_{11}$ -- all but the lowest frequencies disappear, with a few smaller amplitude frequencies present at some orientations but no clear branches. LaZn$_{11}$ does not appear to show the multitude of high ($>4\,\mathrm{kT}$) frequencies seen in CeZn$_{11}$,  probably because its measurement was conducted at lower fields where higher frequencies are less visible due to the damping effect of scattering processes \cite{Shoenberg1984a}. Previously reported dHvA measurements on LaZn$_{11}$ \cite{Hodovanets2013a} -- conducted up to $7\,\mathrm{T}$ at ${\bf B} \parallel [1\,1\,0]$ and ${\bf B} \parallel [0\,0\,1]$ -- observed the same low frequency $\alpha$ and $\beta$ peaks (and the smaller peak at $\sim$$700\,\mathrm{T}$ at ${\bf B} \parallel [0\,0\,1]$) reported here but not the higher frequency features, likely due to the smaller fields used for measurement.

\section{Cyclotron effective masses}
\begin{figure}[h]
\centering
    \includegraphics[width=0.5\textwidth]{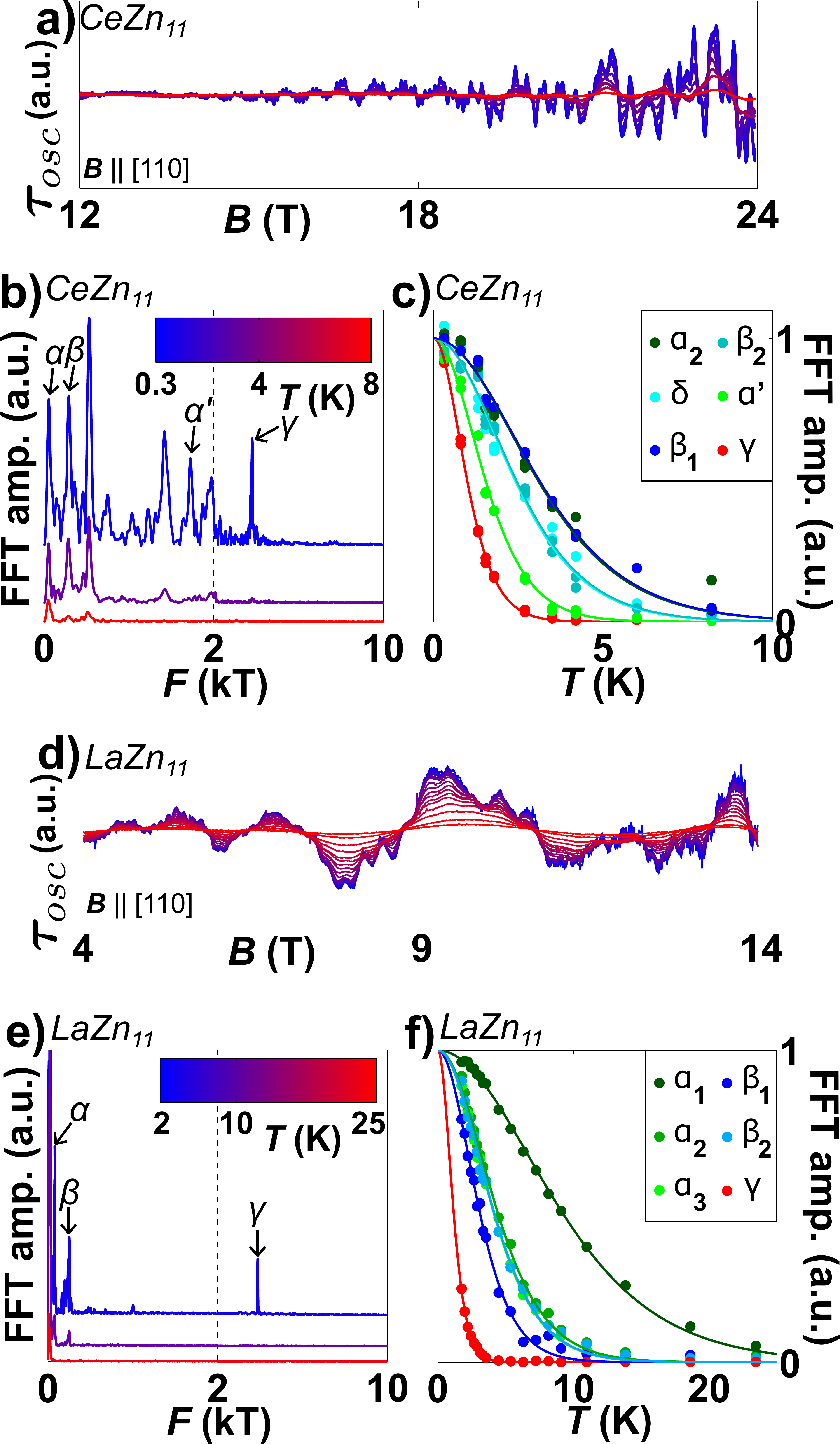}
  \caption{{\bf Temperature dependence of dHvA oscillations in CeZn$_{11}$ and LaZn$_{11}$.} {\bf CeZn$_{11}$}: {\bf a)} Oscillatory component of torque measured close to ${\bf B} \parallel [1\,1\,0]$ between $0.3$ - $8\,\mathrm{K}$ and {\bf b)} the corresponding FFTs for the highest, median and lowest temperatures. The colour of the lines in a) represent temperature on the scale shown in inset of b). {\bf c)} LK fits to the temperature dependence of the oscillation amplitude for different frequencies. {\bf LaZn$_{11}$}: {\bf d)} - {\bf f)} The same three figures, for measurements close to ${\bf B} \parallel [1\,1\,0]$ between $2$ - $25\,\mathrm{K}$. LK amplitudes are scaled at the lowest
  temperature for direct comparison.}
\label{MassData}
\end{figure}

To compare the cyclotron effective masses, $m^*$, of the charge carriers in CeZn$_{11}$ and LaZn$_{11}$, the torque was measured close to ${\bf B} \parallel [1\,1\,0]$ over a range of temperatures for both samples.
 Figures \ref{MassData}a) and d) show the temperature dependence of dHvA oscillations in CeZn$_{11}$ and LaZn$_{11}$ respectively, with the corresponding FFTs shown in Figures \ref{MassData}b) and e). The temperature dependence of amplitudes associated to different dHvA oscillations frequencies  allow us to
extract the value for the cyclotron effective masses, $m^*$,  using the Lifshitz-Kosevich (LK) formula
($R_T=X/\mathrm{sinh}(X)$, where $X=2 \pi^2 k_b T m^* / e \hbar B$  \cite{Shoenberg1984a,Lifshitz1956}). LK fits to the temperature dependence of the amplitudes of the three sets of frequencies $\alpha$, $\beta$ and $\gamma$ are shown in Figures \ref{MassData}c) and f), with the extracted effective masses (and the corresponding frequencies) listed in Table \ref{MassTable}. The frequencies of the $\alpha$, $\beta$ and $\gamma$ peaks are comparable between LaZn$_{11}$ and CeZn$_{11}$, but the effective masses in CeZn$_{11}$ are consistently larger -- taking values between $0.7$ - $2.4\, m_e$ versus $0.1$ - $1.4\, m_e$ in LaZn$_{11}$. This mass enhancement, compared to LaZn$_{11}$, of the effective masses by a factor of $2$ - $4$ in CeZn$_{11}$ suggesting stronger electronic correlations and agrees with the enhancement in the Sommerfeld coefficient of $\sim 3$ obtained from specific heat measurements \cite{Hodovanets2013a}.

In addition to the three sets of common frequencies $\alpha$, $\beta$ and $\gamma$ seen in both CeZn$_{11}$ and LaZn$_{11}$, there are also additional, unassigned frequencies present only in CeZn$_{11}$: a large peak around $500\,\mathrm{T}$ and a number of frequencies between $1$ - $2\,\mathrm{kT}$, in addition to the high ($>4\,\mathrm{kT}$) frequencies seen away from ${\bf B} \parallel [1\,1\,0]$ in Figure \ref{RawData}e). Although the relatively small magnetic fields (up to $14\,\mathrm{T}$ versus up to $33\,\mathrm{T}$ for CeZn$_{11}$) used for the measurement of LaZn$_{11}$ may explain why the high frequencies are not observed, these smaller ($<2\,\mathrm{kT}$) frequencies are of comparable size, amplitude and effective mass (see also Appendix II) to the common frequencies $\alpha$, $\beta$ and $\gamma$, so their absence in LaZn$_{11}$ cannot be due solely to the smaller fields and higher temperatures used for measurement. Furthermore the values (and effective masses) of these unassigned frequencies do not correspond to any multiples of the common frequencies, so they do not originate from any higher order harmonics and their origin is discussed in detail later.

\begin{table*}[t]
\setlength{\tabcolsep}{1.5mm}
\centering
\caption{{\bf Frequencies and effective masses in CeZn$_{11}$ and LaZn$_{11}$}. Comparison between the values found experimentally and those obtained from DFT+$U$ and DFT calculations, when the magnetic field orientation was ${\bf B} \parallel [1\,1\,0]$.}
\hspace*{-4.5mm}
\begin{tabular}{@{ }c@{ } c  c@{ } c c@{ } c  c@{ } c c@{ } c}
\hline
\hline
\noalign{\smallskip}
\multicolumn{3}{l}{${\bf B} \parallel [1 \, 1 \, 0]$} & \multicolumn{3}{l}{{\bf CeZn$_{11}$}} & \multicolumn{4}{c}{{\bf LaZn$_{11}$}} \\
\multicolumn{2}{c}{} & \multicolumn{2}{c}{Experiment} & \multicolumn{2}{c}{DFT+$U$} & \multicolumn{2}{c}{Experiment} & \multicolumn{2}{c}{DFT}\\
\hline
\noalign{\smallskip}
\multicolumn{2}{c}{} & $F$ & $m^*$ & $F$ & $m_b$ & $F$ & $m^*$ & $F$ & $m_b$\\
\multicolumn{1}{c}{} & \multicolumn{1}{@{}c@{}}{Band} & $(\mathrm{T})$ & $(m_e)$ & $(\mathrm{T})$ & $(m_e)$ & $(\mathrm{T})$ & $(m_e)$ & $(\mathrm{T})$ & $(m_e)$\\
\hline
\noalign{\smallskip}
$\beta_1$ & 1 & 260(10) & 0.76(2) & 251 & 0.21 & 220(5) & 0.31(1) & 216 & 0.29 \\
$\beta_2$ & 1 & 300(10) & 0.97(3) & 297 & 0.18 & 250(5) & 0.23(1) & 266 & 0.21 \\
\hline
\noalign{\smallskip}
$\alpha_1$ & 2 & - & - & 35 & 0.09 & 25(5) & 0.10(1) & 25 & 0.10 \\
$\alpha_2$ & 2 & 55(5) & 0.77(3) & 55 & 0.14 & 60(5) & 0.22(1) & 58 & 0.12 \\
$\alpha_3$ & 2 & - & - & 72 & 0.11 & 80(5) & 0.23(1) & 69 & 0.28 \\
$\alpha'$ & 2 & 1730(10) & 1.55(4) & 1516 & 0.82 &  &  &  &  \\
\hline
\noalign{\smallskip}
$\gamma$ & 3 & 3880(10) & 2.39(2) & 4087 & 1.37 & 3880(10) & 1.34(3) & 4237 & 1.18 \\
\hline
\noalign{\smallskip}
- & 4 & - & - & 136 & 2.39 & - & - & 242 & 0.36 \\
$\delta$ & 4 & 170(10) & 0.98(3) & 217 & 0.75 & & & &  \\
\hline
\hline
\end{tabular}
\label{MassTable}
\end{table*}

\section{Density functional theory calculations}
\begin{figure*}[t]
\centering
    \includegraphics[width=1\textwidth]{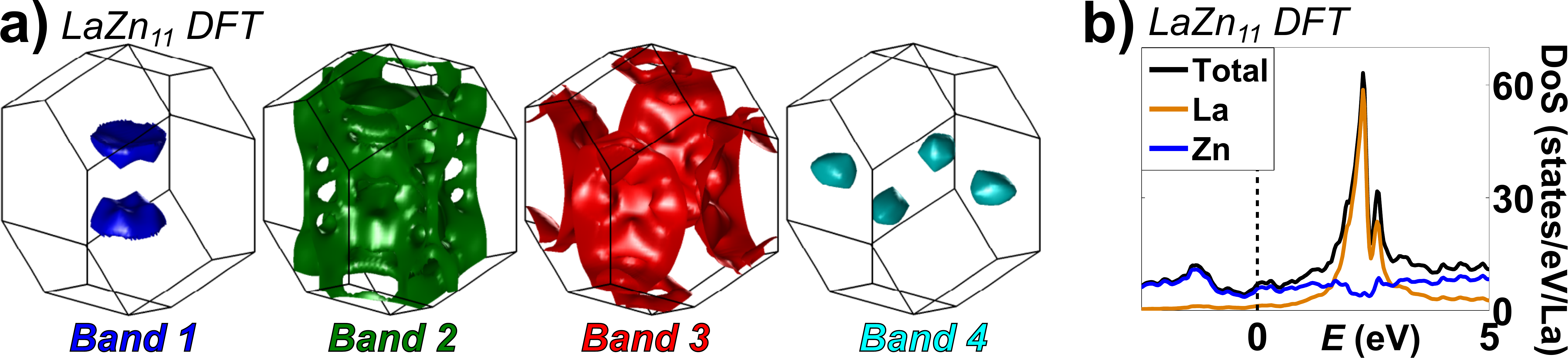}
  \caption{{\bf DFT calculations for LaZn$_{11}$}. {\bf a)} The calculated Fermi surface and {\bf b)} the calculated density of states plot of LaZn$_{11}$.}
\label{FermiSurfacesLaZn11}
\end{figure*}

In order to determine the Fermi surface topology of CeZn$_{11}$ and LaZn$_{11}$, and the origin of the unassigned frequencies seen in CeZn$_{11}$, DFT (and DFT+$U$) calculations were conducted for both LaZn$_{11}$ and CeZn$_{11}$. The DFT calculated Fermi surface of LaZn$_{11}$ is shown in Figure \ref{FermiSurfacesLaZn11}a) and the density of states (DoS) at the Fermi level in Figure \ref{FermiSurfacesLaZn11}b). As La has an unoccupied $4f$ orbital, all of the $4f$ electron states are well above the Fermi level and form the large peak of La states seen close to $2\,\mathrm{eV}$ in the DoS plot of Figure \ref{FermiSurfacesLaZn11}b). Instead the Fermi level is dominated by $3d$ states, which form the continuum of Zn states across the plotted range. The Fermi surface that results from these $3d$ states consists of the four bands shown in Figure \ref{FermiSurfacesLaZn11}a): two small pockets, band 1 (in blue) and band 4 (in cyan); two large and complex surfaces, band 2 (in green) and band 3 (in red).

\begin{figure*}[t]
\centering
    \includegraphics[width=1\textwidth]{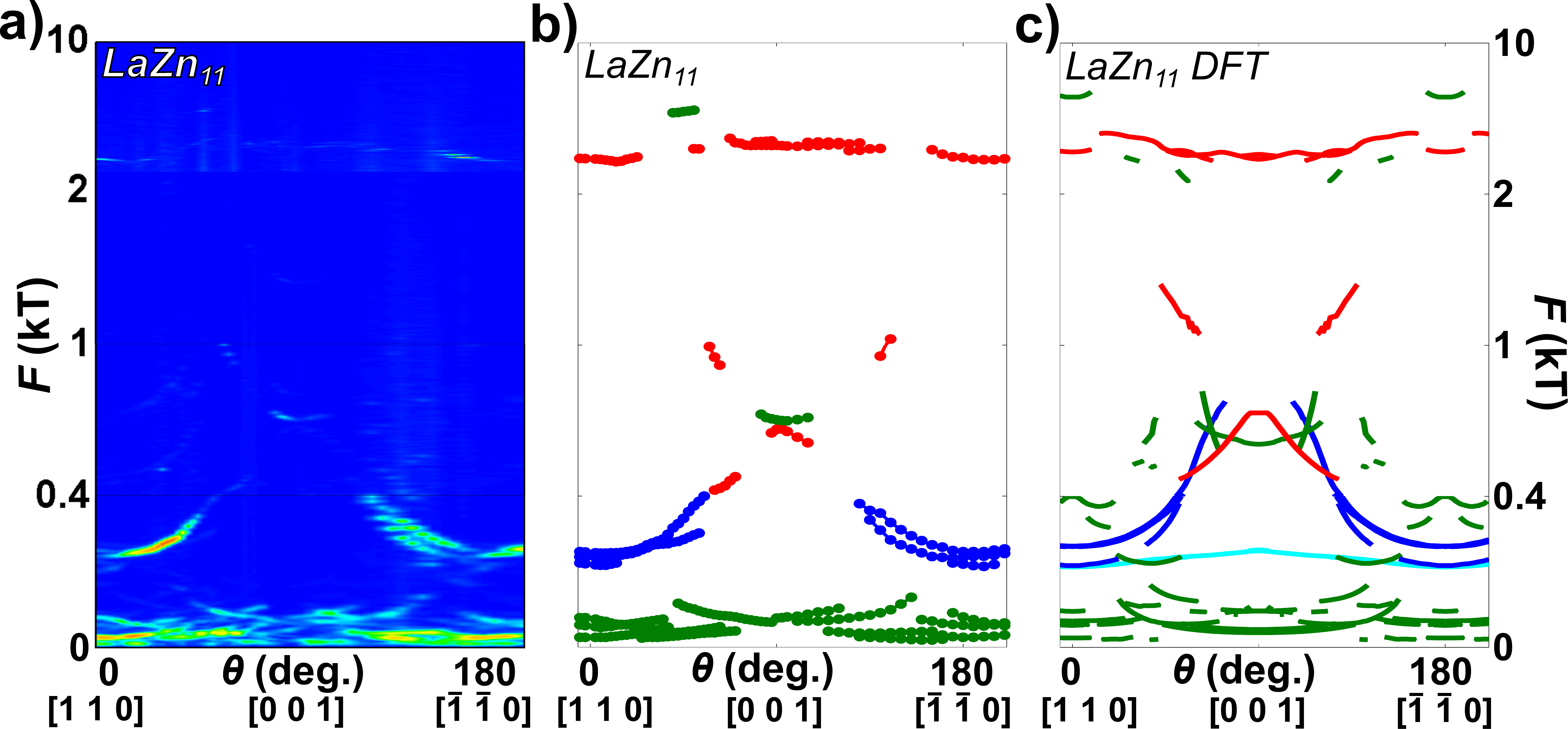}
  \caption{{\bf  Experimental and calculated dHvA frequencies in LaZn$_{11}$.} {\bf a)}  Angular dependence of FFTs showing both the
  frequencies and amplitudes (blue being the background of small amplitude) and {\bf b)} extracted experimental frequencies at $2\,\mathrm{K}$ for angles in the plane containing ${\bf B} \parallel [1\,1\,0]$ ($0^\mathrm{o}$) and ${\bf B} \parallel [0\,0\,1]$ ($90^\mathrm{o}$), alongside  {\bf c)} DFT calculated frequencies}
\label{AngleCompLaZn11}
\end{figure*}

To compare this Fermi surface with the measured dHvA oscillations in LaZn$_{11}$, the frequencies arising from its extremal orbits were calculated over the measured angular range ${\bf B} \parallel [1\,1\,0] \rightarrow {\bf B} \parallel [1\,0\,0]$ and are plotted in Figure \ref{AngleCompLaZn11}c). Figure \ref{AngleCompLaZn11}a) shows the angular dependence of the measured torque FFTs over the same range, with the FFT amplitude determining the colour scale; the dHvA frequencies thus correspond to the peaks in intensity (blue means low amplitude and red high amplitude), and the extracted values are plotted in Figure \ref{AngleCompLaZn11}b). There is very good agreement between the measured and calculated frequencies, with all observed peaks accounted for by the DFT Fermi surface: the set of frequencies $\alpha$ corresponds to orbits around the many small perforations and protrusions of band 2, $\beta$ corresponds to orbits around the pocket of band 1, $\gamma$ corresponds to large orbits traversing the girth of band 3. Additionally, the weaker features seen in smaller regions of the angular range arise from either larger orbits around much of band 2 or smaller orbits around parts of band 3. An exact comparison of the frequencies and masses at ${\bf B} \parallel [1\,1\,0]$ is given in Table \ref{MassTable} -- as well as the agreement in frequency, the effective masses are also very close to the calculated band masses for LaZn$_{11}$. Band 4 is not seen experimentally however, and the reason for this may lie with the value of its
effective masses which can be guessed from its band masses: at ${\bf B} \parallel [1\,1\,0]$ its band mass is close to double that of the comparable sized frequencies of band 1, and away from ${\bf B} \parallel [1\,1\,0]$  its band mass is up to 5 times larger (see Appendix II). These larger band masses for band 4 will lead to greater thermal damping of its dHvA frequencies, to the extent that it may not be visible in the measured temperature range. Other than this, the measured Fermi surface deviates little from that produced by DFT calculations however, so LaZn$_{11}$, with its $4f$ electron states located well above the Fermi level, provides a close facsimile to the Fermi surface expected for CeZn$_{11}$ if the $4f$ electron states are well localised, with Zn $3d$ states dominating at the Fermi level.

\begin{figure*}[t]
\centering
    \includegraphics[width=\textwidth]{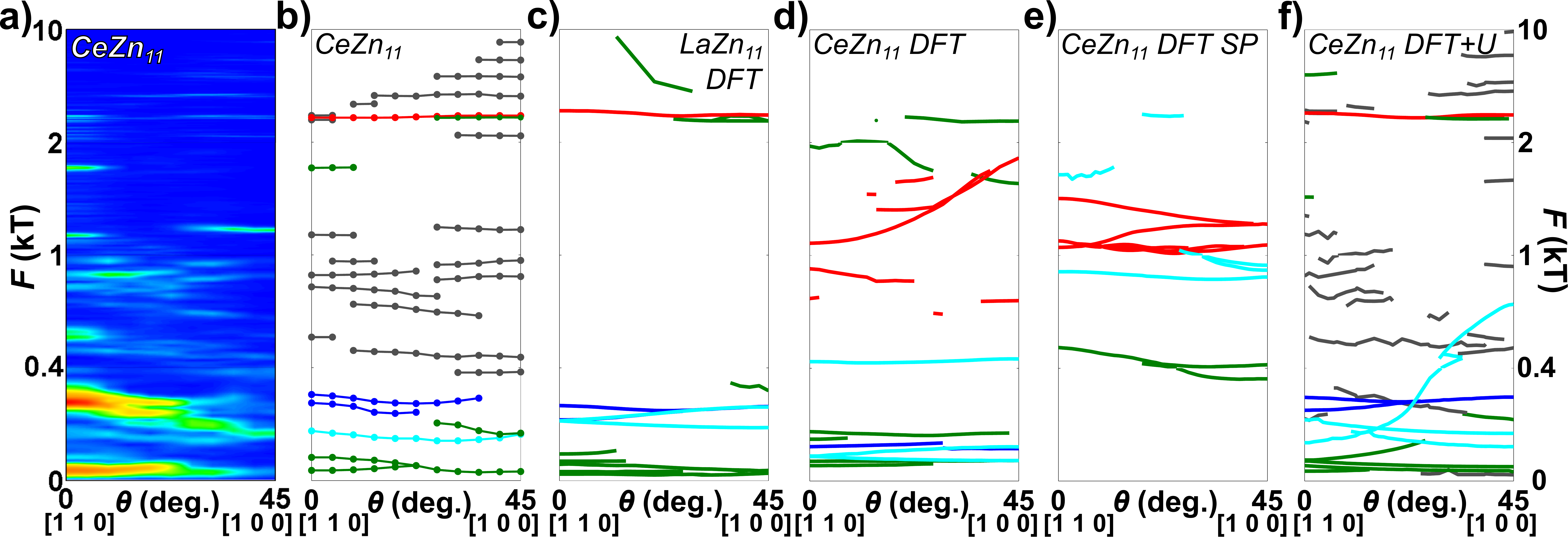}
  \caption{{\bf Experimental and calculated dHvA frequencies in CeZn$_{11}$.} {\bf a)} Angular dependence of FFTs showing both the
  frequencies and amplitudes (blue being the background of small amplitude) and {\bf b)} extracted  experimental frequencies at $320\,\mathrm{mK}$ for angles between ${\bf B} \parallel [1\,1\,0]$ ($0^\mathrm{o}$) and ${\bf B} \parallel [1\,0\,0]$ ($45^\mathrm{o}$), alongside calculated frequencies for {\bf c)} DFT calculations of LaZn$_{11}$, {\bf d)} DFT calculations of CeZn$_{11}$, {\bf e)} spin polarised DFT calculations of CeZn$_{11}$, {\bf f)} DFT+$U$ calculations of CeZn$_{11}$ including magnetic breakdown effects.}
\label{AngleCompCeZn11}
\end{figure*}

The angular dependence of the measured torque FFTs and the extracted dHvA frequencies for CeZn$_{11}$ are shown in Figure \ref{AngleCompCeZn11}a) and b) respectively. Although the three sets of frequencies $\alpha$, $\beta$ and $\gamma$ are still the strongest features, there are many unassigned features at both intermediate ($0.4$ - $1\,\mathrm{kT}$) and high ($>4\,\mathrm{kT}$) frequencies. The frequencies arising from the DFT calculated Fermi surface of LaZn$_{11}$ over the same angular range are plotted in Figure \ref{AngleCompCeZn11}c). As was the case in LaZn$_{11}$, $\alpha$, $\beta$ and $\gamma$ are well matched by orbits of band 2, band 1 and band 3 respectively, although the frequencies of $\beta$ are this time slightly underestimated by the DFT calculations. No features of the LaZn$_{11}$ Fermi surface are able to reproduce the unassigned frequencies seen only in CeZn$_{11}$ however, and the band masses calculated are considerably smaller than the effective masses measured in CeZn$_{11}$.

\begin{figure*}[t]
\centering
    \includegraphics[width=0.95\textwidth]{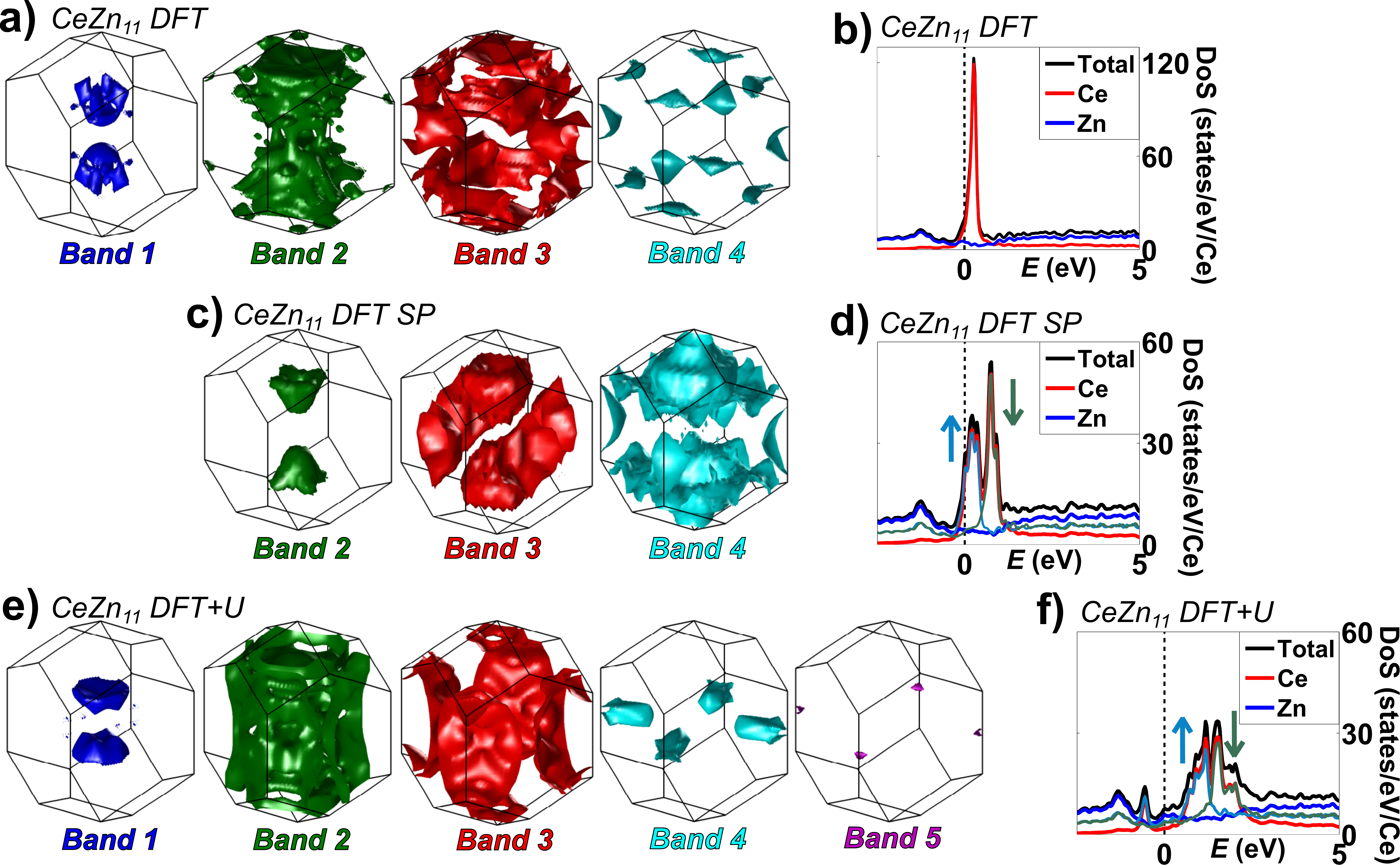}
  \caption{{\bf DFT calculations for CeZn$_{11}$ }. Fermi surfaces and density of state plots are shown for the following calculations: {\bf a)} and {\bf b)} DFT calculations, {\bf c)} and {\bf d)} Spin polarised DFT calculations ($U$=0), {\bf e)} and {\bf f)} DFT+$U$ calculations of CeZn$_{11}$. The different colours used to represent the Fermi surface are used later to represent the calculated dHvA frequencies corresponding to different bands. }
\label{FermiSurfacesCeZn11}
\end{figure*}

Figure \ref{FermiSurfacesCeZn11}a) shows the DFT calculated Fermi surface for CeZn$_{11}$. Although this Fermi surface still consists of four broadly similar bands to LaZn$_{11}$ -- the small pockets of bands 1 (blue) and 4 (cyan), the large and complex surfaces of bands 3 (green) and 4 (red) -- its exact topology is considerably different to that of LaZn$_{11}$, with a larger total area. The reason for this is clear from the DoS plot of Figure \ref{FermiSurfacesCeZn11}b): there is a large peak of Ce $4f$ states located right at the Fermi level, with the $1/14$th of the peak beneath the Fermi level corresponding to the single $4f$ electron of the Ce atom. The Fermi surface thus has a significant contribution from these $4f$ states -- as well as the continuum of Zn $3d$ states -- which acts to substantially change the possible Fermi surface orbits. Figure \ref{AngleCompCeZn11}d) shows the frequencies produced by the DFT calculated CeZn$_{11}$ Fermi surface over the measured angular range. Although band 2 is considerably changed from the case of LaZn$_{11}$, there are still many small orbits which may correspond to the set of frequencies $\alpha$. However the frequencies of band 1 are significantly reduced and no longer agree well with the measured values of $\beta$, while no large orbits of band 3 are a possible match with the measured frequency $\gamma$. The DFT calculation of CeZn$_{11}$ is therefore unable to reproduce the measured dHvA frequencies.

To remove the $4f$ states from the Fermi level DFT+$U$ calculations were performed for CeZn$_{11}$. These calculations introduce a Hubbard term to account for the Coulomb repulsion between electrons occupying the $4f$ states, with $U$ the energy penalty for a doubly occupied state \cite{Anisimov1991}. This acts to lower the energy states of one spin configuration and raise those of the other and so, with a large enough $U$, one can localise the occupied $4f$ electron states well beneath the Fermi level. DFT+$U$ calculations must therefore be spin polarised in order to treat the two spin configurations separately, and this spin polarisation can also affect the electronic structure, with the introduction of a spin-dependent term in the exchange-correlation potential \cite{VonBarth1972}.

The Fermi surface produced by a spin polarised DFT calculation (effectively DFT+$U$ with  $U=0\,\mathrm{eV}$) of CeZn$_{11}$ is shown in Figure \ref{FermiSurfacesCeZn11}c) and the DoS in Figure \ref{FermiSurfacesCeZn11}d). In the DoS the large Ce $4f$ peak has been split into two: a spin down peak (in green) raised in energy, and a spin up peak (in blue) which still lies right at the Fermi level. The resultant Fermi surface now consists of only three bands, with the decrease in energy of the spin up states acting to lower band 1 entirely below the Fermi level, shrink bands 2 and 3 in size and greatly expands band 4. As a result, the orbits associated with this Fermi surface, shown in Figure \ref{AngleCompCeZn11}e) for the measured angular range, are of relatively similar size, falling between $400\,\mathrm{T}$ and $1.5\,\mathrm{kT}$. Although this range corresponds to that of the weaker intermediate dHvA frequencies, no parts of the spin polarised Fermi surface are able to generate frequencies close to the three clearest features, $\alpha$, $\beta$ and $\gamma$, seen in measurements. The spin polarised DFT calculation is therefore also unable to reproduce the measured dHvA frequencies.

 To localise the Ce $4f$ electron, DFT+$U$ calculations were performed for CeZn$_{11}$. A value of $U=1 \, \mathrm{eV}$ was found to be insufficient to move the $4f$ states away from the Fermi level, with a Fermi surface identical to that of the $U=0$ calculation in Figure \ref{FermiSurfacesCeZn11}c). However a relatively modest value of $U=1.5 \, \mathrm{eV}$ was able to localise the $4f$ electron, despite being smaller than the values of $U=4$ - $5\,\mathrm{eV}$ typically applied to Ce-based heavy fermion materials \cite{Dong2014}. The resulting Fermi surface and DoS plot shown in Figure \ref{FermiSurfacesCeZn11}e) and f) respectively.  In the DoS, the spin down states have still been raised in energy and the spin up states lowered, but the spin up peak no longer resides at the Fermi level -- instead the large $U$ has acted to localise a single spin up $4f$ electron, creating the small Ce peak beneath the Fermi level, and raise in energy the remaining spin up states away from the Fermi level. The Fermi level is thus dominated by Zn $3d$ states, as was the case in LaZn$_{11}$, with only a small Ce $4f$ contribution. Unsurprisingly the Fermi surface looks much like that of LaZn$_{11}$, with bands 1 - 3 near identical. Band 4, however, is extended in-plane, so that the four pockets now join at the edge of the Brillouin zone to form a cross-shaped Fermi surface, while an additional band, band 5 (in purple), has appeared, albeit consisting of a single pocket too small to be observed in dHvA measurements.

The frequencies arising from this DFT+$U$ calculated Fermi surface of CeZn$_{11}$, shown as the coloured lines in Figure \ref{AngleCompCeZn11}f), are very similar to those for DFT calculations of LaZn$_{11}$ over the same angular range, shown in Figure \ref{AngleCompCeZn11}c). Band 1 has increased very slightly in size while band 3 has decreased, as can be seen by the frequency values given in Table \ref{MassTable}. The DFT calculated orbits still agree very well with the observed frequency peaks of $\alpha$, $\beta$ and $\gamma$, and a slight increase in the size of band 1 would bring the frequencies of $\beta$ into even closer agreement. The slight changes to band 2 also result in the appearance of a larger, $1.5\,\mathrm{kT}$ frequency close to ${\bf B} \parallel [1\,1\,0]$, and this frequency is labelled $\alpha'$ in  Figure \ref{MassData}b). The largest changes occur to the frequencies of band 4 where the orbits are no longer around small pockets but instead around a larger, cross-shaped Fermi surface. The associated band masses are still much larger than those of the other bands, however for one of the frequency branches  -- that which is largest at ${\bf B} \parallel [1\,1\,0]$ and decreases with rotation towards ${\bf B} \parallel [1\,0\,0]$ -- they are relatively smaller than the rest ($0.75 \, m_e$ versus $2.39 \, m_e$ in Table \ref{MassTable}). This branch can be seen weakly in the measured FFTs of Figure \ref{AngleCompCeZn11}a) and is extracted and plotted in Figure \ref{AngleCompCeZn11}b), labelled $\delta$ in Table I. For bands 1 - 3 however the band masses of the DFT+$U$ calculated Fermi surface of CeZn$_{11}$ are comparable to those found in LaZn$_{11}$ (both in calculation and experiment) and not the larger effective masses measured in CeZn$_{11}$, as can be seen in Table \ref{MassTable}. This confirms that the presence of Ce in CeZn$_{11}$ leads to a definite mass enhancement, in contrast to LaZn$_{11}$ where any mass enhancement above the calculated band mass is negligible.

\section{Magnetic breakdown orbits}
\begin{figure}[h]
\centering
    \includegraphics[width=0.45\textwidth]{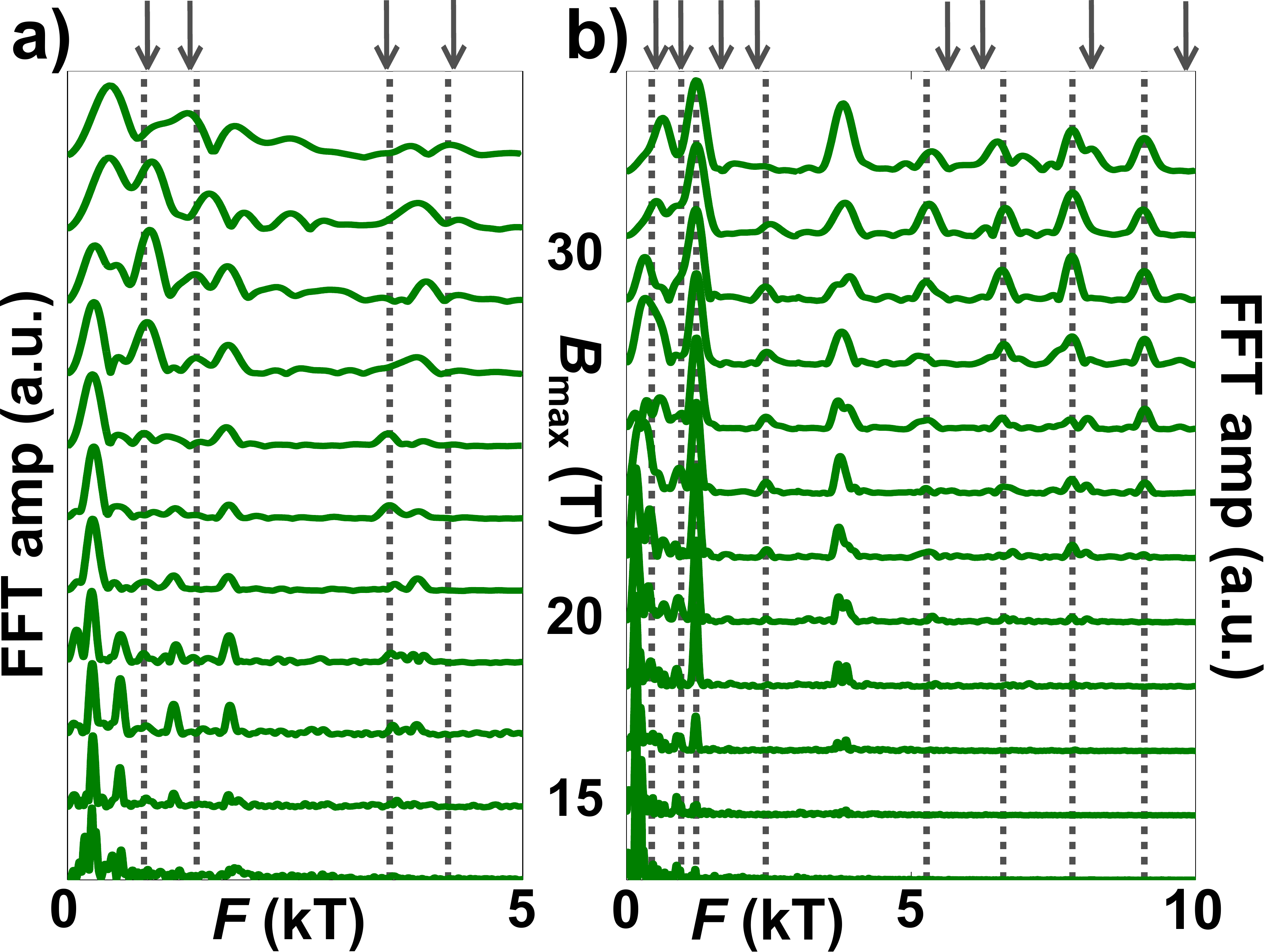}
  \caption{{\bf Field dependence of FFTs in CeZn$_{11}$} at {\bf a)} ${\bf B} \parallel [1\,1\,0]$ and {\bf b)} ${\bf B} \parallel [1\,0\,0]$, performed over $5\,\mathrm{T}$ field windows, $B_{max}$, between $6$ - $33\,\mathrm{T}$. Different spectra for the different field windows are shifted for clarity.
   Dashed lines and arrows indicate the positions of the measured and DFT+$U$ calculated magnetic breakdown frequencies, respectively.}
\label{BreakdownFFTs}
\end{figure}

Although the DFT+$U$ calculated Fermi surface of CeZn$_{11}$ is able to match the strongest measured frequency peaks of $\alpha$, $\beta$ and $\gamma$, its orbits cannot account for the weaker, unassigned intermediate and high frequencies. Figures \ref{BreakdownFFTs}a) and b) show the magnetic field dependence of the FFTs at ${\bf B} \parallel [1\,1\,0]$ and ${\bf B} \parallel [1\,0\,0]$ respectively, with the positions of the unassigned frequency peaks marked by dashed grey lines. At both orientations it can be seen that the unassigned peaks onset at higher fields than the nearby $\alpha$, $\beta$ or $\gamma$ peaks, which is unexpected for frequencies of similar size and effective mass originating from (presumably) the same bands. However, this is the behaviour expected in the case of magnetic breakdown where, at high fields, charge carriers are able to tunnel between adjacent bands of the Fermi surface if their separation in reciprocal space is small enough. This process, depending on the complexity of the Fermi surface and the number of tunnelling points, can generate many new frequencies that may be observed as dHvA oscillations \cite{Carter2010}. The higher fields needed for magnetic breakdown thus explain why breakdown frequencies are observed in CeZn$_{11}$ (measured up to $33\,\mathrm{T}$) and not LaZn$_{11}$ (measured up to $14\,\mathrm{T}$). Explicitly the field dependence is contained within the tunnelling probability $P=\mathrm{exp} (- B_{BD} / B)$, where the breakdown field $B_{BD}=(\pi \hbar / 2e)(k_g^3/(a+b))^{1/2}$ can be considered the field above which the breakdown orbit may be observed; $k_g$ is the tunnelling gap and $1/a$ and $1/b$ are the radii of curvature of the two bands at the tunnelling point \cite{Chambers1966}.

As the DFT+$U$ calculated Fermi surface of CeZn$_{11}$ is so complex there are a multitude of potential tunnelling points located throughout the Brillouin zone, capable of connecting all bands of the Fermi surface indefinitely across reciprocal space. By applying a number of caveats to rule out the least probable orbits, as shown in detail in the Supplementary Material, the possible magnetic breakdown orbits arising from CeZn$_{11}$ were calculated over the measured angular range. The grey arrows in Figure~\ref{BreakdownFFTs}a) and b) indicate the frequencies of the calculated magnetic breakdown orbits for the two high symmetry orientations ${\bf B} \parallel [1\,1\,0]$ and ${\bf B} \parallel [1\,0\,0]$: their frequency values correspond well with the unassigned frequency peaks (indicated by the dashed grey lines), especially at ${\bf B} \parallel [1\,1\,0]$. The breakdown fields of these orbits take values between $B_{BD}=22$ - $39\,\mathrm{T}$, comparable to the measured field range, with the exception of the higher frequencies at ${\bf B} \parallel [1\,0\,0]$ which take larger values of $B_{BD}=120$ - $290\,\mathrm{T}$. These breakdown frequencies show the most deviation from the measured values, thus an adjustment to the shape of these orbits to bring the frequencies into better agreement may also decrease the band separation, and hence the breakdown field, to a smaller value closer to the measured range. The magnetic breakdown frequencies across the measured angular range are shown in Figure \ref{AngleCompCeZn11}f), plotted in dark grey beneath the rest of the coloured frequencies arising from conventional orbits of the CeZn$_{11}$ DFT+$U$ calculated Fermi surface. We find a good correspondence in both size and number with the unassigned experimental dHvA frequencies of CeZn$_{11}$ (dark grey points in Figure \ref{AngleCompCeZn11}b))
with those extracted from the calculated magnetic breakdown orbits in Figure \ref{AngleCompCeZn11}f). Some discrepancies that still exist, in both the size of the frequencies and the angular range over which they are visible, may arise from slight differences between the experimental and  DFT+$U$ calculated  CeZn$_{11}$ Fermi surfaces as a slight adjustment to the orbit shape may decrease the breakdown field, expanding the angular range over which the frequency is visible. Nevertheless, the consideration of magnetic breakdown orbits means that the DFT+$U$ calculated Fermi surface of CeZn$_{11}$ is able to fully describe the dHvA oscillations seen in CeZn$_{11}$ (see also Appendix I), and so its Fermi surface looks much like that of Figure \ref{FermiSurfacesCeZn11}e), with the occupied $4f$ electron states localised well below the Fermi level.

In conclusion, we report a dHvA study of the Fermi surface of the antiferromagnet CeZn$_{11}$, alongside its non-magnetic analogue LaZn$_{11}$, to determine the contribution of the $4f$ Ce electrons to its electronic and magnetic properties. Torque measurements detect a field-induced magnetic phase transition in CeZn$_{11}$, but both CeZn$_{11}$ and LaZn$_{11}$ show similar dHvA frequencies originating from a complex, multi-band Fermi surface. We compare our results with DFT and DFT+$U$ calculations and find that the measured frequencies of LaZn$_{11}$ are well described by a four band Fermi surface, dominated by Zn $3d$ states. For CeZn$_{11}$ both DFT and spin polarised DFT calculations place a large peak of Ce $4f$ states at the Fermi level and produce Fermi surfaces that do not agree with the measured dHvA frequencies. However DFT+$U$ calculations with $U$=1.5 $\mathrm{eV}$ produce a Fermi surface very similar to that found in experiments and localise the occupied $4f$ states well below the Fermi level. Additional, weaker frequencies observed only in CeZn$_{11}$ are found to originate from magnetic breakdown orbits, which can occur at higher fields where tunnelling between orbits becomes possible. The cyclotron effective masses are a factor of $2$ - $4$ larger in CeZn$_{11}$ than in LaZn$_{11}$. Our study confirms that the occupied $4f$ electron states are well localised in CeZn$_{11}$, resulting in a Fermi surface close to that of LaZn$_{11}$. However the larger cyclotron effective masses found in CeZn$_{11}$, compared to LaZn$_{11}$, suggest the localised $4f$ electrons are still able to enhance the electronic correlations, as well as determining the magnetic properties of the system.

\section{Acknowledgements}
We thank Roser Valenti and Milan Tomic for useful discussions and Arjun Narayanan and Matthew Watson for technical support. This work was mainly supported by the EPSRC (EP/L001772/1, EP/I004475/1, EP/I017836/1). Part of this work was supported supported by
HFML-RU/FOM, member of the European Magnetic Field  Laboratory (EMFL) and by EPSRC (UK) via its membership to the EMFL
(grant no. EP/N01085X/1). The authors would like to acknowledge the use of the University of Oxford Advanced Research Computing (ARC) facility in carrying out part of this work. Work done at Ames Lab was supported by the U.S. Department of Energy,
Office of Basic Energy Science, Division of Materials Sciences and Engineering. Ames Laboratory is operated for the U.S. Department of Energy by Iowa State University under Contract No. DE-AC02-07CH11358. AIC acknowledges an EPSRC Career Acceleration Fellowship (EP/I004475/1).

\bibliography{CeZn11paper}

\newpage
\newpage

\vspace{30cm}

\section{Appendix I. Calculation of magnetic breakdown orbits}

\begin{figure}[h]
\centering
    \includegraphics[width=0.48\textwidth]{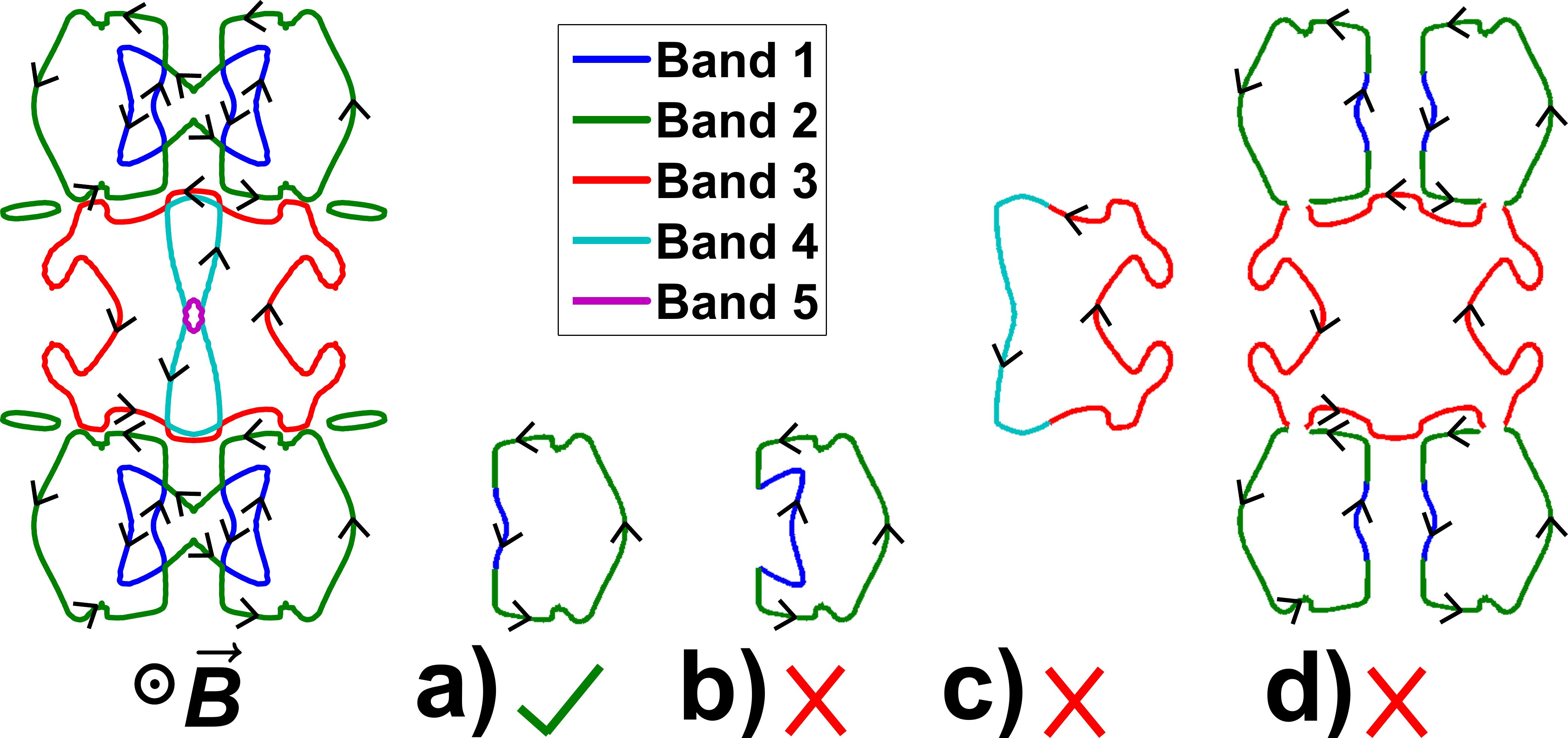}
  \caption{{\bf Possible and impossible magnetic breakdown orbits in CeZn$_{11}$}, shown for a slice centered on $[0\,0\,0]$ for the field orientation ${\bf B} \parallel [1\,0\,0]$. Arrows denote the direction of electronic motion for the field direction indicated.}
\label{CeZn11_allowedorbits}
\end{figure}

As the the DFT+U calculated Fermi surface of CeZn$_{11}$ is rather complicated, there are a multitude of potenetial magnetic breakdown tunnelling points located throughout the Brillouin zone.
Calculating an infinite number of orbits is obviously impossible, and even an intermediate number of orbits will be computationally intensive -- each additional tunnelling point exponentially increases the number of orbits -- and gratuitous -- only a small number of additional orbits are seen experimentally. However further constraints can be placed on
the orbits calculated, these are shown pictorially in Figure \ref{CeZn11_allowedorbits}. In an applied field the electrons will all process in the same direction (i.e. clockwise or anti-clockwise, depending on the geometry) around the fundamental orbits, and this must also be the case for the magnetic breakdown orbits.

 Therefore the orbit of Figure \ref{CeZn11_allowedorbits}a) is allowed -- the electrons travel in the same direction for all constituent parts of the orbit -- but the orbit of Figure \ref{CeZn11_allowedorbits}b) is not. Furthermore band 4 is unlikely to contribute to any observable magnetic breakdown orbits: its fundamental frequencies are for the most part not observed due to their comparatively larger masses, and any magnetic breakdown orbits arising from these frequencies would be of even smaller amplitude. For this reason the orbit of Figure \ref{CeZn11_allowedorbits}c) is unlikely to be experimentally observable, and so all similar orbits involving band 4 are discounted. Finally, the probability of an orbit occuring decreases with every tunnelling point passed, whether or not a tunnelling event occurs. Therefore orbits encompassing a large number of tunnelling points, such as that shown in Figure \ref{CeZn11_allowedorbits}d), are highly improbable, and the amplitudes of the associated frequencies will be vanishingly small; these can also be neglected in calculations.

\begin{figure*}[t]
\centering
    \includegraphics[width=\textwidth]{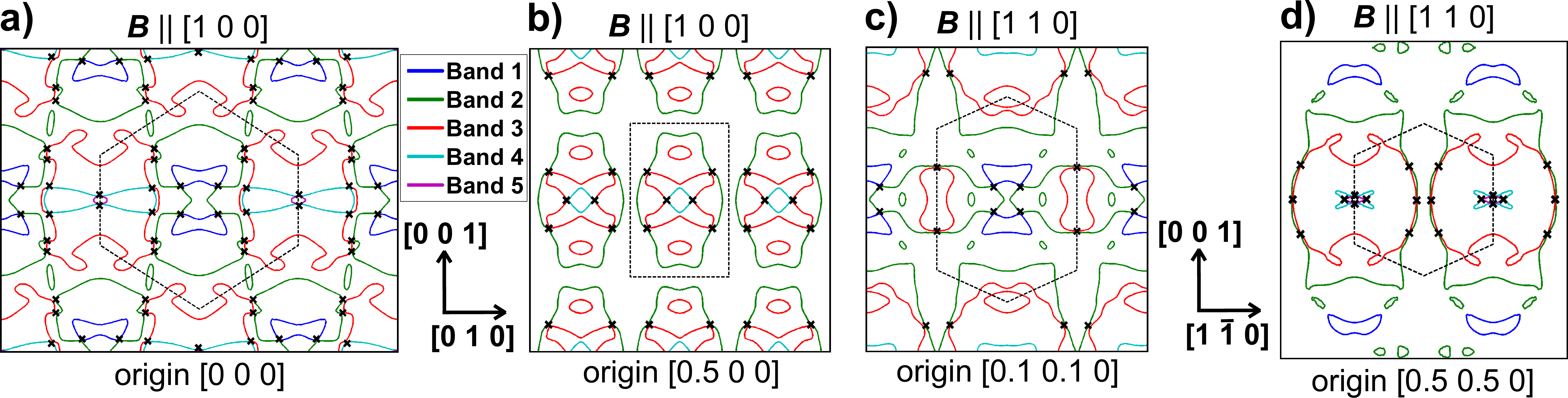}
 \caption{{\bf Slices of the Fermi surface of CeZn$_{11}$} for the orientations ${\bf B} \parallel [1\,0\,0]$ centred on {\bf a)} $[0\,0\,0]$ and {\bf b)} $[0.5\,0\,0]$, and ${\bf B} \parallel [1\,1\,0]$ centred on {\bf c)} $[0.1\,0.1\,0]$ and {\bf d)} $[0.5\,0.5\,0]$. Points where tunnelling between bands may be possible are marked with an \textbf{x}. Centre coordinates are given in terms of the reciprocal lattice vectors.}
 \label{CeZn11_slices}
 \vspace{2cm}
    \includegraphics[width=\textwidth]{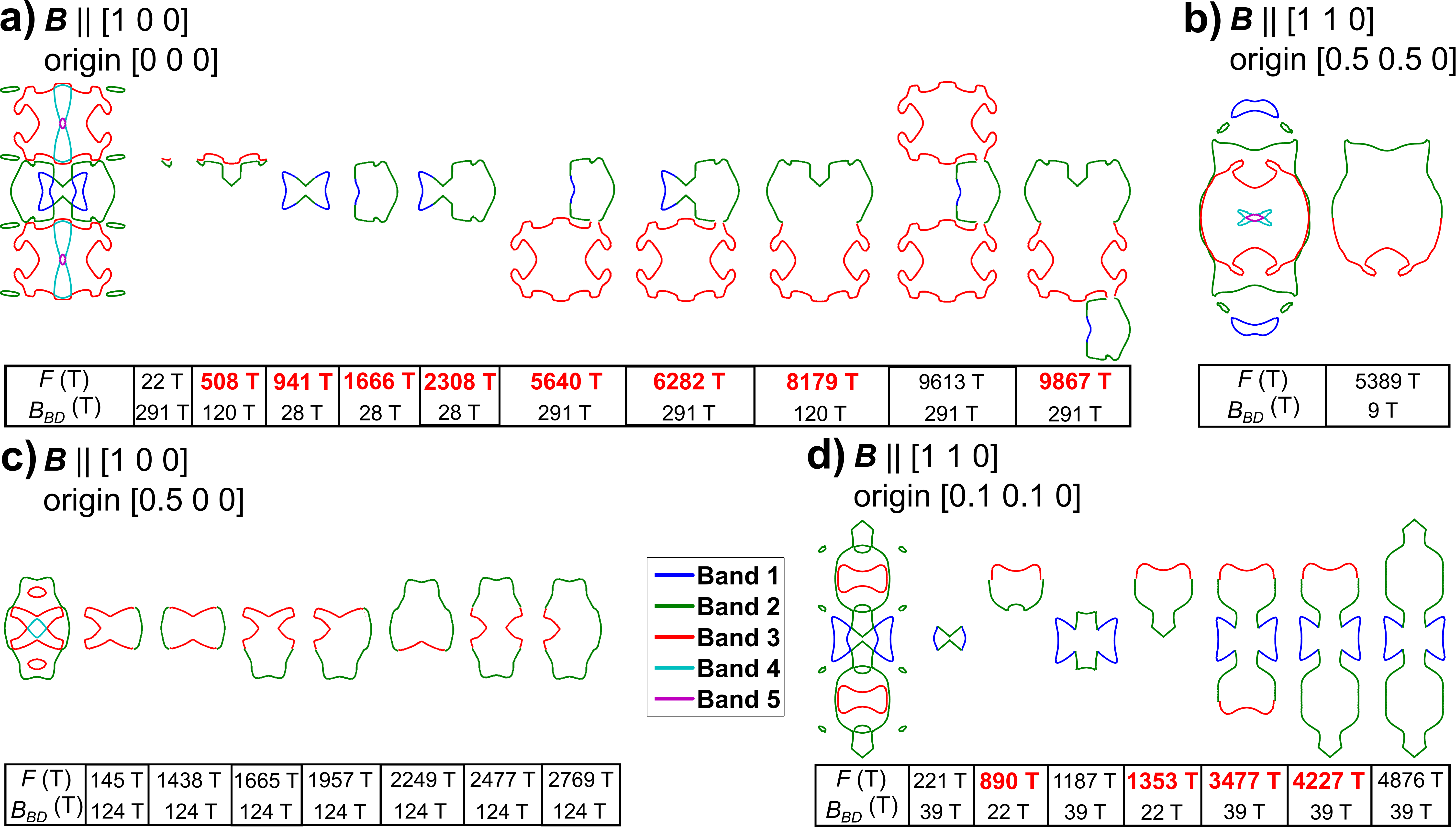}
    \caption{{\bf Calculated magnetic breakdown orbits in CeZn$_{11}$}. The original Fermi surface (far left) and calculated breakdown orbits for the orientations ${\bf B} \parallel [1\,0\,0]$ centred on {\bf a)} $[0\,0\,0]$ and {\bf c)} $[0.5\,0\,0]$, and ${\bf B} \parallel [1\,1\,0]$ centred on {\bf b)} $[0.5\,0.5\,0]$ and {\bf d)} $[0.1\,0.1\,0]$. Red, bolded frequencies are seen experimentally.}
\label{CeZn11_breakdownorbits}
\end{figure*}

Figure \ref{CeZn11_slices} shows slices of the Fermi surface normal to the high symmetry directions $[1\,0\,0]$ and $[1\,1\,0]$, where the bands are in closest proximity. Points where tunnelling may be possible between bands are marked with an \textbf{x}. In the simplest cases, such as slice d), the tunnelling points link two orbits of the Fermi surface that are well separated from the rest. The magnetic breakdown of these orbits will lead to two or three additional frequencies, of comparable size to the parent orbits. However a more complicated case is presented in slice c), where tunnelling via band 1 is able to link orbits of band 2 in adjacent Brillouin zones, allowing for the construction of a potentially infinite number of possible orbits. In slice a), where all parts of the Fermi surface are interconnected by tunnelling points, things are even more complicated still.

Figures \ref{CeZn11_breakdownorbits}a),c) and b),d) show the magnetic breakdown orbits calculated, with the prior constraints applied, for the orientations ${\bf B} \parallel [1\,0\,0]$ and ${\bf B} \parallel [1\,1\,0]$ respectively. The frequency values and associated breakdown fields are given beneath, with the orbits seen experimentally marked in red and bolded. Most of the orbits found have reassuringly low breakdown fields, in the range $22\,\mathrm{T}<B_{BD}<39\,\mathrm{T}$, compared to a maximum measured field of $33\,\mathrm{T}$. The exceptions to this are the orbits between band 2 and band 3 shown in Figure \ref{CeZn11_breakdownorbits}a), which take values of $B_{BD}=120\,\mathrm{T}$ and $291\,\mathrm{T}$ for the outer and inner tunnelling points respectively. These same orbits also showed the largest discrepancy with their measured frequency values, so it is likely that in reality bands 2 and 3 take a slightly modified form at this orientation, both lowering $B_{BD}$ and better matching the experimental frequencies.


Not all the calculated breakdown orbits are seen, most notably those shown in Figure \ref{CeZn11_breakdownorbits}b) and c). In the latter case the high breakdown field, $B_{BD}=124\,\mathrm{T}$, is likely prohibitive; in the former case the high frequency ($5389\,\mathrm{T}$) and effective mass ($1.12\,m_e$) will damp the amplitude of oscillation. The few frequencies in Figure \ref{CeZn11_breakdownorbits}a) and d) that are not clearly seen experimentally all occur in the proximity of other frequencies, either of fundamental or breakdown origin, making them difficult to distinguish.

\newpage
\section{Appendix II. Other temperature dependence of dHvA oscillations in LaZn$_{11}$ and CeZn$_{11}$.}

This section present additional figures of the temperature dependence of dHvA oscillations in LaZn$_{11}$ and CeZn$_{11}$.
The results are compared with the results of DFT calculations for different field orientations and the extracted parameters
are listed in the Tables below. Possible magnetic breakdown orbits for the CeZn$_{11}$ are also listed and they can explain the additional high frequencies of weak amplitude observed for CeZn$_{11}$ as compared with LaZn$_{11}$.

\begin{figure}[h!]
\centering
\includegraphics[width=0.45\textwidth]{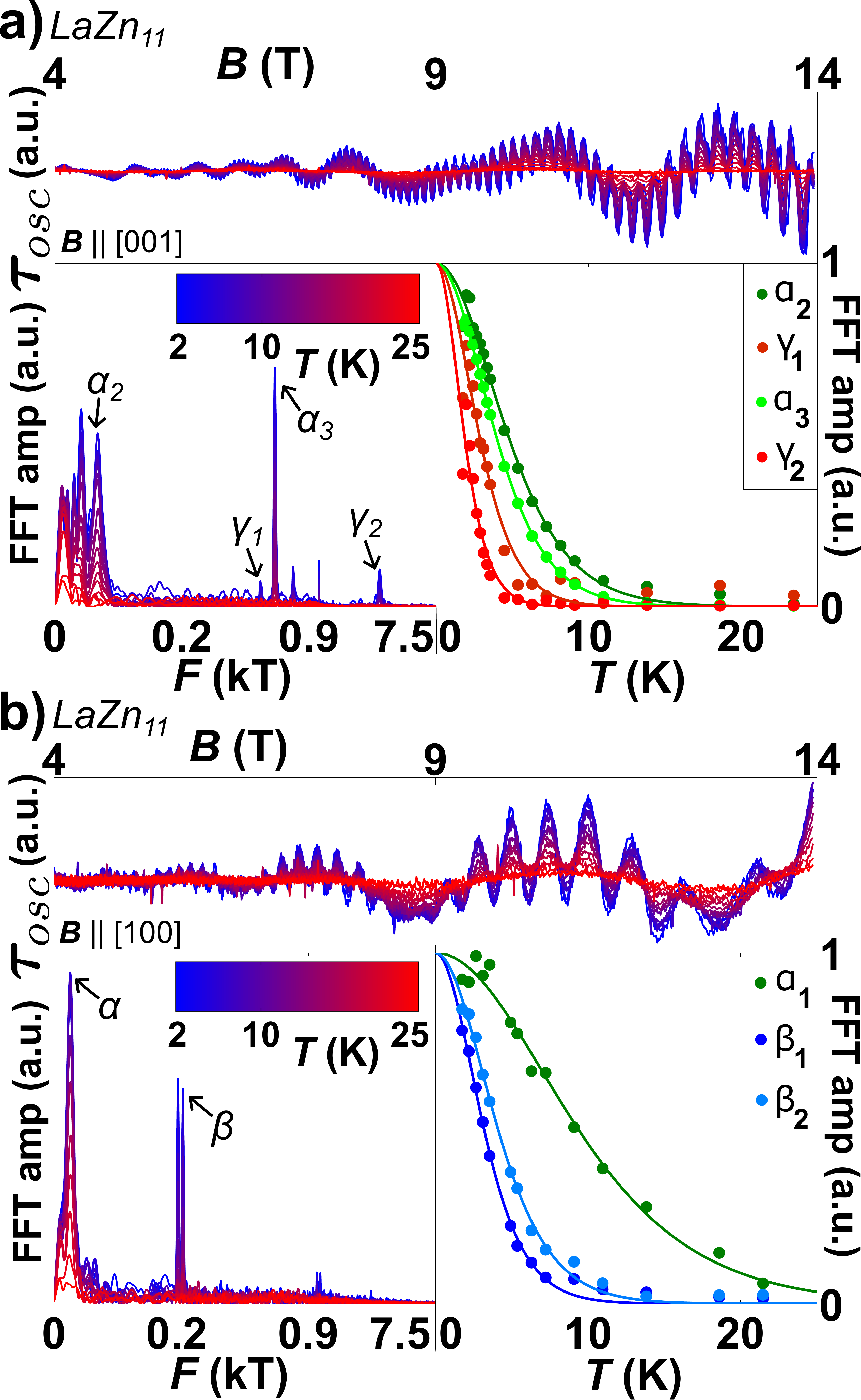}
  \caption{{\bf Further temperature dependence of dHvA oscillations in LaZn$_{11}$.} (Top) Oscillatory component of torque, (bottom left) the corresponding FFTs and (bottom right) LK fits to the temperature dependence of the oscillation amplitude for the labelled peaks, for measurements close to {\bf a)} ${\bf B} \parallel [0\,0\,1]$ and {\bf b)} ${\bf B} \parallel [1\,0\,0]$ between $2$ - $25\,\mathrm{K}$.}
 \label{LaZn11_temp}
\end{figure}

\begin{figure}[h!]
\centering
    \includegraphics[width=0.45\textwidth]{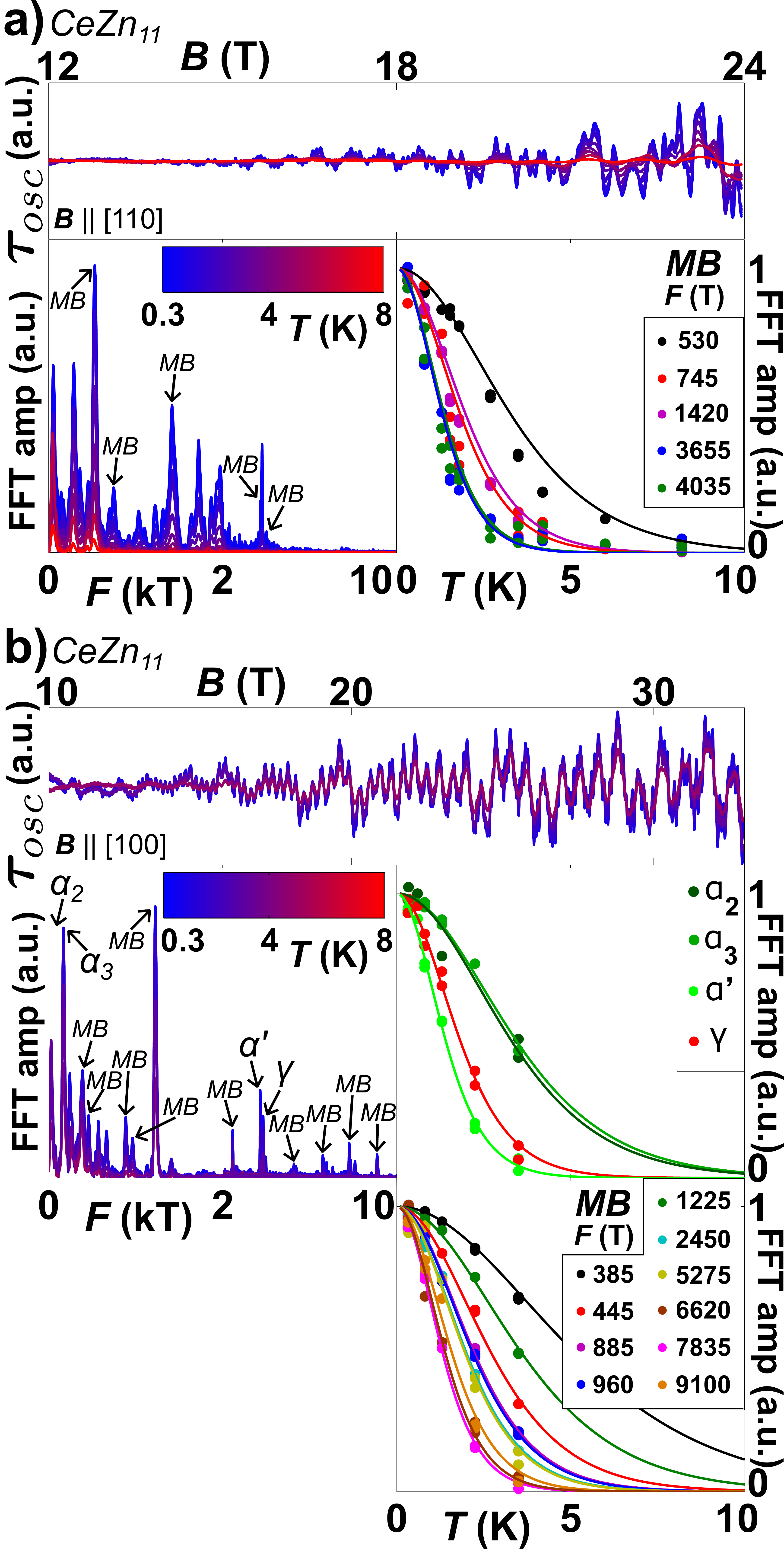}
  \caption{{\bf Further temperature dependence of dHvA oscillations in CeZn$_{11}$.} (Top) Oscillatory component of torque, (bottom left) the corresponding FFTs and (bottom right) LK fits to the temperature dependence of the oscillation amplitude for the labelled peaks, for measurements close to {\bf a)} ${\bf B} \parallel [1\,1\,0]$, with only the magnetic breakdown peaks shown, and {\bf b)} ${\bf B} \parallel [1\,0\,0]$, with the conventional orbits and magnetic breakdown orbits shown, between $0.3$ - $8\,\mathrm{K}$.}
\label{CeZn11_temp}
\end{figure}

\newpage

\begin{table*}[ht!]
\setlength{\tabcolsep}{1.5mm}
\centering
\caption{{\bf Frequencies and effective masses in LaZn$_{11}$} comparing those found experimentally with the results of DFT calculations, for the field orientations ${\bf B} \parallel [0\,0\,1]$ and ${\bf B} \parallel [1\,0\,0]$.}
\label{LaZn11_MassTable}
\begin{tabular}{@{ }c@{ } c  c@{ } c c@{ } c  c@{ } c c@{ } c}
\hline
\hline
\noalign{\smallskip}
\multicolumn{3}{l}{{\bf LaZn$_{11}$}} & \multicolumn{3}{l}{${\bf B} \parallel [0 \, 0 \, 1]$} & \multicolumn{4}{c}{${\bf B} \parallel [1 \, 0 \, 0]$} \\
\multicolumn{2}{c}{} & \multicolumn{2}{c}{Experiment} & \multicolumn{2}{c}{DFT} & \multicolumn{2}{c}{Experiment} & \multicolumn{2}{c}{DFT}\\
\hline
\noalign{\smallskip}
\multicolumn{2}{c}{} & $F$ & $m^*$ & $F$ & $m_b$ & $F$ & $m^*$ & $F$ & $m_b$\\
\multicolumn{1}{c}{} & \multicolumn{1}{@{}c@{}}{Band} & $(\mathrm{T})$ & $(m_e)$ & $(\mathrm{T})$ & $(m_e)$ & $(\mathrm{T})$ & $(m_e)$ & $(\mathrm{T})$ & $(m_e)$\\
\hline
\noalign{\smallskip}
$\beta_1$ & 1 &  &  &  &  & 205(5) & 0.29(1) & - & - \\
$\beta_2$ & 1 &  &  &  &  & 235(5) & 0.23(1) & 263 & 0.15 \\
\hline
\noalign{\smallskip}
$\alpha_1$ & 2 & - & - & 39 & 0.18 & 25(5) & 0.10(1) & 36 & 0.11 \\
$\alpha_2$ & 2 & 65(5) & 0.20(1) & 47 & 0.23 & - & - & 320 & 0.33 \\
 & 2 & - & - & 3849 & 0.29 & - & - & 3499 & 0.81 \\
$\alpha_3$ & 2 & 710(5) & 0.38(1) & 606 & 0.20 &  &  &  &  \\
\hline
\noalign{\smallskip}
$\gamma_1$ & 3 & 635(10) & 0.53(4) & 731 & 0.56 &  &  &  &  \\
 & 3 & - & - & 3849 & 0.29 & - & - & 3930 & 1.58 \\
$\gamma_2$ & 3 & 3880(10) & 2.39(2) & 4087 & 1.37 &  &  &  &  \\
\hline
\noalign{\smallskip}
- & 4 &  &  &  &  & - & - & 188 & 0.27 \\
- & 4 & - & - & 255 & 0.82 & - & - & 261 & 0.75 \\
\hline
\hline
\end{tabular}
\end{table*}

\begin{table*}[htbp]
\setlength{\tabcolsep}{1.5mm}
\centering
\caption{{\bf Frequencies and effective masses in CeZn$_{11}$} comparing those found experimentally with the results of DFT+U calculations including magnetic breakdown, for the field orientations ${\bf B} \parallel [1\,1\,0]$ and ${\bf B} \parallel [1\,0\,0]$.}
\begin{tabular}{@{ }c@{ } c  c@{ } c c@{ } c  c@{ } c c@{ } c}
\hline
\hline
\noalign{\smallskip}
\multicolumn{3}{l}{{\bf CeZn$_{11}$}} & \multicolumn{3}{l}{${\bf B} \parallel [1 \, 1 \, 0]$} & \multicolumn{4}{c}{${\bf B} \parallel [1 \, 0 \, 0]$} \\
\multicolumn{2}{c}{} & \multicolumn{2}{c}{Experiment} & \multicolumn{2}{c}{DFT+U} & \multicolumn{2}{c}{Experiment} & \multicolumn{2}{c}{DFT+U}\\
\hline
\noalign{\smallskip}
\multicolumn{2}{c}{} & $F$ & $m^*$ & $F$ & $m_b$ & $F$ & $m^*$ & $F$ & $m_b$\\
\multicolumn{1}{c}{} & \multicolumn{1}{@{}c@{}}{Band} & $(\mathrm{T})$ & $(m_e)$ & $(\mathrm{T})$ & $(m_e)$ & $(\mathrm{T})$ & $(m_e)$ & $(\mathrm{T})$ & $(m_e)$\\
\hline
\noalign{\smallskip}
$\beta_1$ & 1 & 260(10) & 0.76(2) & 251 & 0.21 &  &  &  &  \\
$\beta_2$ & 1 & 300(10) & 0.97(3) & 297 & 0.18 & - & - & 298 & 0.12 \\
\hline
\noalign{\smallskip}
$\alpha_1$ & 2 & - & - & 35 & 0.09 & - & - & 32 & 0.08 \\
$\alpha_2$ & 2 & 55(5) & 0.77(3) & 55 & 0.14 & 50(10) & 0.71(4) & 51 & 0.10 \\
$\alpha_3$ & 2 & - & - & 72 & 0.11 & 170(5) & 0.68(2) & 218 & 0.29 \\
$\alpha'$ & 2 & 1730(10) & 1.55(4) & 1516 & 0.82 & 3730(10) & 1.66(5) & 3683 & 0.72 \\
\hline
\noalign{\smallskip}
$\gamma$ & 3 & 3880(10) & 2.39(2) & 4087 & 1.37 & 3865(10) & 1.28(4) & 3953 & 1.38 \\
\hline
\noalign{\smallskip}
- & 4 & - & - & 136 & 2.39 & - & - & 122 & 1.38 \\
$\delta$ & 4 & 170(10) & 0.98(3) & 217 & 0.75 & 170(5) & 0.68(2) & 169 & 0.27 \\
- & 4 &  &  &  &  & - & - & 740 & 1.93 \\
\hline
Magnetic & breakdown  & orbits &  &  &  &  &  &  &  \\
\noalign{\smallskip}
\hline
\multicolumn{2}{c|}{\it MB} & 530(10) & 0.95(3) & - & - & 385(10) & 0.61(2) & - & - \\
\multicolumn{2}{c|}{\it MB} & 745(10) & 1.7(1) & 893 & 0.84 & 445(5) & 1.15(3) & 508 & 0.02 \\
\multicolumn{2}{c|}{\it MB} & 1420(10) & 1.57(3) & 1353 & 0.86 & 885(10) & 1.40(2) & - & - \\
\multicolumn{2}{c|}{\it MB} & 3655(10) & 2.36(9) & 3477 & 1.34 & 960(10) & 1.43(2) & 941 & 0.38 \\
\multicolumn{2}{c|}{\it MB} & 4035(10) & 2.29(9) & 4227 & 1.46 & 1225(5) & 0.87(2) & 1666 & 0.48 \\
\multicolumn{2}{c|}{\it MB} &  &  &  &  & 2450(10) & 1.56(3) & 2308 & 0.66 \\
\multicolumn{2}{c|}{\it MB} &  &  &  &  & 5275(10) & 1.59(8) & 5640 & 1.13 \\
\multicolumn{2}{c|}{\it MB} &  &  &  &  & 6620(10) & 2.24(9) & 6282 & 0.94 \\
\multicolumn{2}{c|}{\it MB} &  &  &  &  & 7835(20) & 2.37(4) & 8179 & 0.68 \\
\multicolumn{2}{c|}{\it MB} &  &  &  &  & 9100(10) & 1.97(9) & 9867 & 0.26 \\
\hline
\hline
\end{tabular}

\label{CeZn11_MassTable}
\end{table*}

\end{document}